\documentclass[a4paper]{jpconf}
\usepackage{graphicx}
\begin{document}
\title{The unquenched quark model \footnote{Talk presented at 38th Symposium on Nuclear Physics, January  6-9 2015, Cocoyoc Morelos (Mexico).}}
\author{H. Garc{\'{\i}}a-Tecocoatzi}
\address{Instituto de Ciencias Nucleares, Universidad Nacional Aut\'onoma de M\'exico, AP 70-543, 04510 M\'exico DF, M\'exico}
\address{Dipartimento di Fisica and INFN, Universit\'a di Genova, via Dodecaneso 33, I-16146 Genova, Italy}

\author{R. Bijker}
\address{Instituto de Ciencias Nucleares, Universidad Nacional Aut\'onoma de M\'exico, AP 70-543, 04510 M\'exico DF, M\'exico}


\ead{hugo.garcia@nucleares.unam.mx}
'
\begin{abstract}
In this contribution, we briefly analyze the formalism of the unquenched quark model (UQM) and its application to the  description of several observables of hadrons. In the UQM, the effects of $q \bar q$ sea pairs are introduced explicitly into the quark model through a QCD-inspired $^{3}P_0$ pair-creation mechanism. We present our description of   flavour asymmetry  and strangeness in the proton when baryon-meson components are included.    In the meson sector,  the charmonium  and bottomonium
 spectra with self-energy corrections due to the coupling to the meson-meson components . \end{abstract}

\section{Introduction}
The quark model \cite{Eichten:1974af,Isgur:1979be,Godfrey:1985xj,Capstick:1986bm,Giannini:2001kb,Glozman-Riska,Loring:2001kx,Ferretti:2011,Galata:2012xt,BIL} can reproduce the behavior of observables such as the spectrum and the magnetic moments in the baryon and meson sector, but it neglects quark-antiquark pair-creation (or continuum-coupling) effects.
Above threshold, these couplings lead to strong decays and  below threshold, they  lead to virtual $q \bar q - q \bar q$ ($qqq - q \bar q$) components in the hadron wave function and shifts of the physical mass with respect to the bare mass. 
The unquenching of the quark model for hadrons is a way to take these components into account.

Pioneering works on the unquenching of quark model were  done by T\"ornqvist and collaborators, who used  an unitarized QM \cite{Ono:1983rd,Tornqvist}, while 
Van Beveren and Rupp used an 
t-matrix approach \cite{vanBeveren:1979bd,vanBeveren:1986ea}.
These methods were used (with a few variations) by several authors to study the influence of the meson-meson (meson-baryon) continuum on meson (baryon) observables. 
These techniques were applied to study of the scalar meson nonet ($a_0$, $f_0$, etc.) of Ref. \cite{vanBeveren:1986ea,Tornqvist:1995kr} in which the loop contributions are given by the hadronic intermediate states that each meson can access. It is via these hadronic loops that the bare states become ``dressed'' and  the hadronic loop contributions totally dominate the dynamics of the process.   On the other hand, Isgur and coworkers in Ref. \cite{Geiger:1989yc} demonstrated that the effects of the $q \bar q$ sea pairs in meson spectroscopy is simply a renormalization of the meson string tension.  Also, the strangeness content of the nucleon and electromagnetic form factors were  investigated in  \cite{Geiger:1996re,Bijker:2012zza}, whereas  Capstick and Morel in Ref. \cite{Capstick} analyzed  
baryon meson loop effects on the spectrum of nonstrange baryons.   
In meson sector, Eichten {\it et al.} explored the influence of the open-charm channels on the charmonium properties using the Cornell coupled-channel model \cite{Eichten:1974af} to assess departures from the single-channel  potential-model expectations.

In this contribution, we discuss some of the latest applications of the UQM (the approach is a generalization of the unitarized quark model \cite{vanBeveren:1979bd,vanBeveren:1986ea,Tornqvist,Tornqvist:1995kr}) to study the flavor asymmetry  and strangeness  of the proton,
 in wich the effects of the quark-antiquark pairs were introduced  into the constituent quark model (CQM) in a systematic way and the wave fuctions were given explicitly.  Finally, the UQM is applied 
 to describe meson observables and the spectroscopy of the charmonium  and bottomonium.	

\section{UQM }
\label{Sec:formalism}
\begin{figure}[h]
\begin{minipage}{14pc}
\includegraphics[width=14pc]{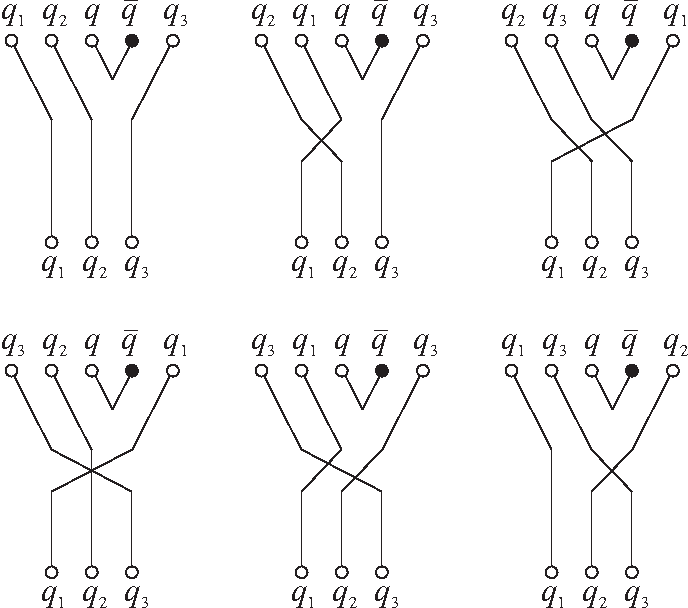}
\caption{\label{figbaryon}Quark line diagrams for $A \rightarrow BC$ with $q \bar{q} = s \bar{s}$ and
$q_1 q_2 q_3 = uud$}
\end{minipage}\hspace{2pc}%
\begin{minipage}{20pc}
\includegraphics[width=20pc]{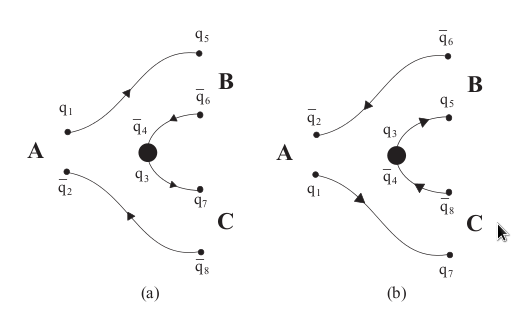}
\caption{\label{figmeson} Two diagrams can contribute to the process $A \rightarrow BC$. $q_i$
and $q_i$ stand for the various initial (i = 1 - 4) and final (i = 5 - 8)
quarks or antiquarks, respectively.}
\end{minipage} 
\end{figure}

In the unquenched quark model for baryons \cite{Bijker:2012zza,Santopinto:2010zza,Bijker:2009up,Bijker:210} and mesons \cite{bottomonium,charmonium,Ferretti:2013vua,Ferretti:2014xqa}, the hadron wave function is made up of a zeroth order $qqq$ ($q \bar q$) configuration plus a sum over the possible higher Fock components, due to the creation of $^{3}P_0$ $q \bar q$ pairs. Thus,  we have 
\begin{eqnarray} 
	\label{eqn:Psi-A}
	\mid \psi_A \rangle &=& {\cal N} \left[ \mid A \rangle 
	+ \sum_{BC \ell J} \int d \vec{K} \, k^2 dk \, \mid BC \ell J;\vec{K} k \rangle \right.
	\nonumber\\
	&& \hspace{2cm} \left.  \frac{ \langle BC \ell J;\vec{K} k \mid T^{\dagger} \mid A \rangle } 
	{E_a - E_b - E_c} \right] ~, 
\end{eqnarray}
where $T^{\dagger}$ stands for the $^{3}P_0$ quark-antiquark pair-creation operator \cite{bottomonium,charmonium,Ferretti:2013vua,Ferretti:2014xqa}, $A$ is the baryon/meson, $B$ and $C$ represent the intermediate state hadrons,  see Figures \ref{figbaryon} and \ref{figmeson}. $E_a$, $E_b$ and $E_c$ are the corresponding energies, $k$ and $\ell$ the relative radial momentum and orbital angular momentum between $B$ and $C$ and $\vec{J} = \vec{J}_b + \vec{J}_c + \vec{\ell}$ is the total angular momentum. 
It is worthwhile noting that in Refs. \cite{bottomonium,charmonium,Ferretti:2013vua,Ferretti:2014xqa,Kalashnikova:2005ui}, the constant pair-creation strength in the operator (\ref{eqn:Psi-A}) was substituted with an effective one, to suppress unphysical heavy quark pair-creation. 

In the UQM 
 the matrix elements of an observable $\hat {\mathcal {O}}$ can be calculated as
\begin{equation}
	O = \left\langle \psi_A \right| \hat{ \mathcal {O} }\left| \psi_A \right\rangle \mbox{ }, 
\end{equation}
where $\left| \psi_A \right\rangle$ is the state of Eq. (\ref{eqn:Psi-A}). 
The result will receive a contribution from the valence part and one from the continuum component, which is absent in naive QM calculations. 
 
The introduction of continuum effects in the QM can thus be essential to study observables that only depend on $q \bar q$ sea pairs, like the strangeness content of the nucleon electromagnetic form factors \cite{Geiger:1996re,Bijker:2012zza} or the flavor asymmetry of the nucleon sea \cite{Santopinto:2010zza}. 
In other cases, continuum effects can provide important corrections to baryon/meson observables, like the self-energy corrections to meson masses \cite{bottomonium,charmonium,Ferretti:2013vua,Ferretti:2014xqa} or the importance of the orbital angular momentum in the spin of the proton \cite{Bijker:2009up}.

\section{Flavour content in the proton}
 
The first  evidence for the flavor asymmetry of the proton sea was provided 
by NMC at CERN \cite{nmc}. The flavor asymmetry in the proton is related to the Gottfried integral 
for the difference of the proton and neutron electromagnetic structure functions 
\begin{eqnarray}
S_G = \int_0^1 dx \frac{F_2^p(x)-F_2^n(x)}{x} 
= \frac{1}{3} - \frac{2}{3} \int_0^1 dx \left[ \bar{d}(x) - \bar{u}(x) \right] ~.
\end{eqnarray}
Under the assumption of a flavor symmetric sea, one obtains the 
Gottfried sum rule $S_G=1/3$. The final NMC value is $0.2281 \pm 0.0065$ at $Q^2 = 4$ 
(GeV/c)$^2$ for the Gottfried integral over the range $0.004 \leq x \leq 0.8$ \cite{nmc}, 
which implies a flavor asymmetric sea. The violation of the Gottfried sum rule has been 
confirmed by other experimental collaborations \cite{hermes,nusea}. 
Theoretically, it was shown  in Ref. \cite{Thomas}, that the coupling of the nucleon to the pion 
cloud provides a natural mechanism to produce a flavor asymmetry. 
\begin{figure}
\begin{center}
\includegraphics[width=6cm]{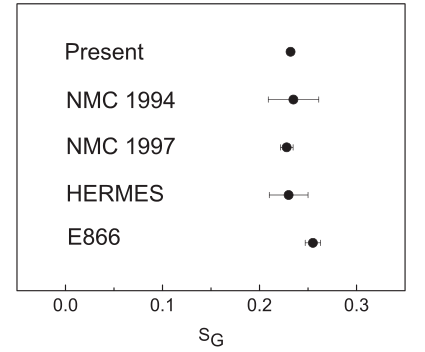}
\caption{\label{protonasy}Comparison the value of Gottfried sum rule calculated within UQM with the experimental data  from NMC 1994, NMC 1997, HERMES, and E866. Figure  taken 
from Ref. \cite{Bijker:210}; APS copyright. }
\end{center}
\end{figure}
In the UQM, the flavor asymmetry can be 
calculated from the difference of the probability to find  $\bar{d}$ and $\bar{u}$ sea quarks in the proton 
\begin{eqnarray}
N_{\bar{d}}-N_{\bar{u}} 
= \int_0^1 dx \left[ \bar{d}(x) - \bar{u}(x) \right] ~. 
\label{asym}
\end{eqnarray}
Note that, even in absence of explicit information on the (anti)quark distribution 
functions, the integrated value can be obtained directly from the left-hand side 
of Eq.~(\ref{asym}).  Our result is shown in  Fig. \ref{protonasy}.

\begin{figure}[h]
\begin{minipage}{18pc}
\includegraphics[width=18pc]{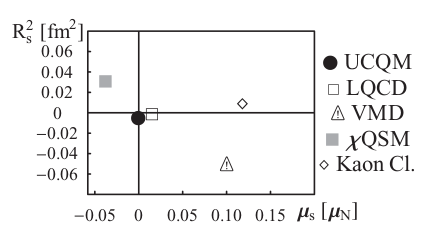}
\caption{\label{smagnetic}TheUQM results for the strange magnetic moment and radius of the proton. Figure taken from Ref. \cite{Bijker:2012zza}; APS copyright.}
\end{minipage}\hspace{2pc}%
\begin{minipage}{15pc}
\includegraphics[width=15pc]{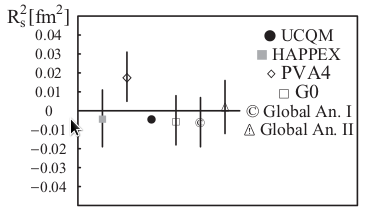}
\caption{\label{sradio} Comparison between our resulting value for the strange radius of the proton in the UQM. Figure  taken from Ref. \cite{Bijker:2012zza};  APS copyright.}
\end{minipage} 
\end{figure}

The results for the two strangeness observables were obtained in a calculation involving a sum over intermediate states up to four oscillator shells for both baryons and mesons
\cite{Bijker:2012zza}. In the UQM formalism, the strange magnetic moment of the proton is defined as the expectation value of the operator
\begin{eqnarray}
\vec{\mu}_{s}=\sum_{i} \mu_{i,s}\left[2\vec{s}(q_{i})+\vec{l}(q_{i})- 2\vec{s}(\bar{q}_{i})-\vec{l}(\bar{q}_{i}) \right]
\end{eqnarray}
on the proton state of Eq. (\ref{eqn:Psi-A}), which represents the contribution of the strange quarks to the magnetic moment fo the proton; $\mu_{i,s}$ is the magnetic moment of the quark i times a projector on strangeness and the strange quark magnetic moment is set as in Ref. \cite{Bijker:210}. Our result  is  $\vec{\mu}_{s}=0.0006 \mu_N$(see Fig.\ref{smagnetic}).

Similarly, the strange radius of the proton is defined as the expectation value of the operator
\begin{eqnarray}
R^{2}_{s}= \sum^{5}_{i=1}e_{i,s}\left(\vec{r}_{i}-\vec{R}_{\rm cm}\right)^2
\end{eqnarray}
on the proton state of Eq. (\ref{eqn:Psi-A}), where $e_{i,s}$ is the electric charge
of the quark $i$ times a projector on strangeness, $\vec{r}_{i}$  and $\vec{R}_{\rm cm}$ are the coordinates of the quark $i$ and of the intermediate state center of mass, respectively.
The expectation value of $R^{2}_{s}$ on the proton is equal to $-0.004 {\rm fm}^2 $. In Fig. \ref{sradio} our result is compared with the experimental data.

\section{Self-energy corrections in the UQM}
In Refs. \cite{bottomonium,charmonium,Ferretti:2013vua,Ferretti:2014xqa}, the method was used by some of us to compute the charmonium  ($c \bar c$) and
bottomonium ($b \bar b$) spectra with self-energy corrections, due to continuum coupling effects. 
In the UQM, the physical mass of a meson, 
\begin{equation}
	\label{eqn:self-trascendental}
	M_a = E_a + \Sigma(E_a)  \mbox{ },
\end{equation}
is given by the sum of two terms: a bare energy, $E_a$, calculated within a potential model \cite{Godfrey:1985xj}, and a self energy correction, 
\begin{equation}
	\label{eqn:self-a}
	\Sigma(E_a) = \sum_{BC\ell J} \int_0^{\infty} k^2 dk \mbox{ } \frac{\left| M_{A \rightarrow BC}(k) \right|^2}{E_a - E_b - E_c}  \mbox{ },
\end{equation}
computed within the UQM formalism. 

\begin{figure}[h]
\begin{minipage}{16pc}
\includegraphics[width=16pc]{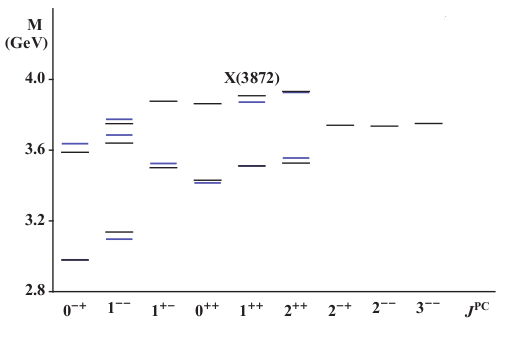}
 \caption{\label{charm} Charmonium spectrum with self energies corrections. 
  Black lines are theoretical predictions and blue lines are experimental data available. Figure taken from Ref. \cite{charmonium}; APS copyright.}\end{minipage}\hspace{2pc}%
\begin{minipage}{18pc}
\includegraphics[width=18pc]{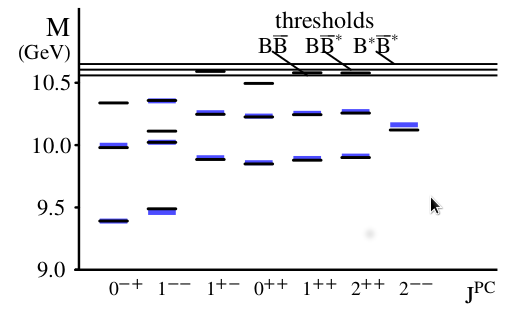}
 \caption{\label{botton}Bottomonium spectrum with self energies corrections.
 Black lines are theoretical predictions and blue lines are experimental data available. Figure taken from Ref. \cite{Ferretti:2013vua}; APS copyright. 
 }

\end{minipage} 
\end{figure}
Our results for the self energies corrections  of charmonia \cite{charmonium,Ferretti:2014xqa} and bottomonia \cite{bottomonium,Ferretti:2013vua,Ferretti:2014xqa} spectrums, 
are shown in  figures \ref{charm} and \ref{botton}.

\section{Discussion and conclusion}
In the baryon sector, our results for    asymmetry and "strangeness" observables, as shown in  Figures  \ref{protonasy}, \ref{smagnetic}  and  \ref{sradio}, are in agreement with the experimental data.
These observables can only be understood  when continuum components in the wave function are included. 

Our results in the meson sector  for the self energies corrections  of charmonium  and bottomonium spectra, see figures \ref{charm} and \ref{botton}, show that the pair-creation effects on the spectrum of heavy mesons are quite small. Specifically for charmonium and bottomonium states, they are of the order of $2 - 6\%$ and $1 \%$, respectively. 
The relative mass shifts, i.e. the difference between the self energies of two meson states, are in the order of a few tens of MeV. 

However, as QM's can predict the meson masses with relatively high precision in the heavy quark sector, even these corrections can become significant.
These results are particularly interesting in the case of states close to an open-flavor decay threshold, like the $X(3872)$ and $\chi_b(3P)$ mesons.
In our picture the $X(3872)$ can be interpreted as a $c \bar c$ core [the $\chi_{c1}(2^3P_1)$], plus higher Fock components due to the coupling to the meson-meson continuum. In Ref. \cite{Ferretti:2014xqa}, we showed that the probability to find the $X(3872)$ in its core or continuum components is approximately $45\%$ and $55\%$, respectively.  

In conclusion, the flavor asymmetry in the  proton can be well described by the UQM. The effects of the continuum  components on the "strangeness" observables of the proton are found to be negligible. Nevertheless, our results are compatible with the latest experimental data and recent lattice calculations. In the meson sector our self energies corrections  for 
 charmonia  and bottomonia  are found to be significant.

\section*{Acknowledgments}

This work is supported in part by PAPIIT-DGAPA, Mexico (grant IN107314) and INFN sezione di Genova .


\section*{References}

\begin{thebibliography}{100}

  
\bibitem{Eichten:1974af}  
  ~Eichten E, ~Gottfried K, ~Kinoshita T, ~B~Kogut J,  ~Lane ~D K and ~-M~Yan T 1975
 {\it Phys.\ Rev.\ Lett.\ }  {\bf 34} 369 ;
~Eichten E, ~Gottfried K , ~Kinoshita T, ~D~Lane K and ~-M~Yan T 1978 
  {\it  Phys.\ Rev.} \ D {\bf 17} 3090; 1980  {\it  Phys.\ Rev.} \ D
  {\bf 21}  203. 
	
\bibitem{Isgur:1979be}
~Isgur N and ~Karl G 1978
   {\it Phys.\ Rev.}\  D {\bf 18} 4187;
   1979 {\it Phys.\ Rev.}\  D
  {\bf 19}, 2653 
  [1981 Erratum-ibid.\  {\bf 23} 817];
 1979  {\it Phys.\ Rev.}\  D  {\bf 20} 1191.
	
\bibitem{Godfrey:1985xj}
  ~Godfrey S and ~Isgur N 1985
  {\it Phys.\ Rev.}\ D {\bf 32} 189.

\bibitem{Capstick:1986bm}
  ~Capstick S and ~Isgur N 1986
   {\it Phys.\ Rev.}\  D {\bf 34} 2809 .
	
\bibitem{Giannini:2001kb}
  ~Ferraris M, ~Giannini M M, ~Pizzo M, ~Santopinto E and ~Tiator L 1995
  {\it Phys.\ Lett.} \ B {\bf 364} 231; 
  ~Santopinto E, ~Iachello F and ~Giannini M M 1998
   {\it Eur.\ Phys.\ J.} \ A {\bf 1} 307 ;
 ~ Santopinto  E and ~Giannini M M 2012
   {\it Phys.\ Rev.}\ C {\bf 86} 065202;	
   Giannini M
 M, Santopinto E 2015 {\it Chin. J. Phys.}  {\bf 53}  020301; 
 Aiello M 1996  {\it et al. Phys.Lett.} B  387, 215; Aiello {\it et al.}  1988{\it
 J. of Phys.} G  {\bf 24} 753; Bijker R, Iachello F,  Santopinto E 1988
 {\it J. of  Phys.} A {\bf 31} 9041; M. De Sanctis  {\it et al.}  2007 {\it Phys. Rev.}  C {\bf 6} 
062201.
	
\bibitem{Glozman-Riska}
  ~Glozman L Y, ~Riska D O 1996
   {\it Phys.\ Rept.}\  {\bf 268} 263;
  ~Glozman L Y, ~Plessas W, ~Varga K, ~Wagenbrunn R F 1998
   {\it Phys.\ Rev.}\  D {\bf 58} 094030.  
  
\bibitem{Loring:2001kx}
~Loring   U, ~Metsch B C and ~Petry H R 2001
  {\it Eur.\ Phys.\ J.} \  A {\bf 10} 395. 
	
\bibitem{Ferretti:2011}
 ~Santopinto  E 2005
   {\it Phys.\ Rev.}\  C {\bf 72} 022201;
  ~Ferretti J, ~Vassallo A and ~Santopinto  E 2011
   {\it Phys.\ Rev.}\ C {\bf 83}, 065204;
  ~De~Sanctis M, ~Ferretti J, ~Santopinto  E and ~Vassallo A
  arXiv:1410.0590.	De Sanctis, M, Ferretti J 2011
Santopinto E  {\it et al. Phys. Rev. } C {\bf 84} 055201. 
	
\bibitem{Galata:2012xt}
~Galata G and ~Santopinto  E 2012
   {\it Phys.\ Rev.}\ C {\bf 86} 045202.	
		
\bibitem{BIL}
 Bijker R, Iachello  F and  LeviatanA 1994
 {\it Ann. Phys.} (N.Y.) {\bf 236} 69; 1996
 {\it Phys. Rev.} C {\bf 54} 1935; 1997
 {\it Phys. Rev. }D {\bf 55} 2862; 2000
 {\it Ann. Phys. } (N.Y.) {\bf 284} 89.
	
	
\bibitem{vanBeveren:1979bd} 
  ~van Beveren E, ~Dullemond C and ~Rupp G 1980
   {\it Phys.\ Rev. }\ D {\bf 21} 772
  [ 1980 Erratum-ibid.\ D {\bf 22} 787].

\bibitem{vanBeveren:1986ea} 
  ~van Beveren E, ~Rijken T~A, K.~Metzger, ~Dullemond C, ~Rupp G and ~Ribeiro J E 1986
   {\it Z.\ Phys.}\ C {\bf 30} 615.
  \bibitem{Ono:1983rd}
  ~Ono S and ~T\"ornqvist N~A 1984
   {\it Z.\ Phys.}\ C {\bf 23} 59;
  ~Heikkila K,  ~Ono S and ~T\"ornqvist N~A 1984
   {\it Phys.\ Rev.} \ D {\bf 29} 110
  [1984 Erratum-ibid.\ {\bf 29} 2136];
  ~Ono S, ~Sanda A ~I and ~T\"ornqvist N A 1986
   {\it Phys.\ Rev.}\ D {\bf 34} 186.  
		
\bibitem{Tornqvist}
  N.~A.~T\"ornqvist and P.~Zenczykowski 1984
   {\it Phys.\ Rev.}\  D {\bf 29} 2139; 1986
   {\it Z.\ Phys.}\  C {\bf 30} 83;
  P.~Zenczykowski 1986
   {\it Annals Phys.}\  {\bf 169} 453.	
	
\bibitem{Tornqvist:1995kr} 
~Tornqvist N A 1995
  {\it Z.\ Phys.}\ C {\bf 68} 647.
\bibitem{Pennington:2002} 
  ~Boglione M and  ~Pennington M R 2002
     {\it Phys.\ Rev.}\ D {\bf 65} 114010.	

	
\bibitem{Geiger:1989yc} 
~Geiger P  and ~Isgur N 1990
     {\it Phys.\ Rev.}\ D {\bf 41} 1595.
	
\bibitem{Geiger:1996re} 
 ~Geiger P  and ~Isgur N 1997
     {\it Phys.\ Rev.}\ D {\bf 55} 299.	
		
\bibitem{Bijker:2012zza}  
  ~Bijker R, ~Ferretti J and ~Santopinto E 2012
     {\it Phys.\ Rev.}\ C {\bf 85} 035204.   
\bibitem{Capstick}
Capstick S and Morel D [nucl-th/0204014].
  
\bibitem{Santopinto:2010zza}
~Santopinto  E and ~Bijker R 2010
   {\it  Phys.\ Rev.}\ C {\bf 82} 062202.     
  
\bibitem{Bijker:2009up}
  ~Santopinto  E and ~Bijker R 2008
    {\it Few Body Syst.}\  {\bf 44} 95.
\bibitem{Bijker:210}  
  ~Bijker R and ~Santopinto E 2009
     {\it Phys.\ Rev.}\ C {\bf 80} 065210. 
		
\bibitem{bottomonium}
  ~Ferretti J, ~Galat\`a  G, ~Santopinto E,and ~Vassallo A 2012
     {\it Phys.\ Rev.} \ C {\bf 86} 015204.  
	
\bibitem{charmonium}  
 ~Ferretti J, ~Galat\`a  G and ~Santopinto E 2013
    {\it Phys.\ Rev.}\ C {\bf 88} 015207.	
	
\bibitem{Ferretti:2013vua}   
~Ferretti J,  and ~Santopinto E 2014
     {\it Phys.\ Rev.}\  D {\bf 90} 094022.	
  
\bibitem{Ferretti:2014xqa} 
~Ferretti J, ~Galat\`a  G and ~Santopinto E 2014
     {\it Phys.\ Rev.} \  D {\bf 90} 054010.	
\bibitem{Kalashnikova:2005ui}
~Kalashnikova Y S 2005
    {\it Phys.\ Rev.}\ D {\bf 72} 034010.		
\bibitem{nmc}
 Amaudruz  P {\it et al.} 1991 {\it Phys. Rev. Lett.} {\bf 66} 2712; 
Arneodo  M {\it et al.} 1997 {\it Nucl. Phys. } B {\bf 487} 3.

\bibitem{hermes}
 Ackerstaff K {\it et al.} 1988 {\it Phys. Rev. Lett.} {\bf 81} 5519. 

\bibitem{nusea}
Towell  R S  {\it et al.} 2001 {\it Phys. Rev. }D {\bf 64} 052002.

\bibitem{Thomas}
Thomas A W 1983 {\it Phys. Lett. }B {\bf 126} 97.

\end{thebibliography}
\end{document}